\documentclass[aps,prb,superscriptaddress,floatfix,twocolumn]{revtex4}
\usepackage[ansinew]{inputenc}
\usepackage{longtable}
\usepackage{xspace}
\usepackage{graphicx}
\graphicspath{{../}}

\newcommand{\invcm}{cm\ensuremath{^{-1}}\xspace}
\newcommand{\NH}{\hbox{NH$_3$}\xspace}
\newcommand{\NHplus}{\hbox{NH$_4^{+}$}\xspace}
\newcommand{\NHmoins}{\hbox{NH$_2^{-}$}\xspace}
\newcommand{\ND}{\hbox{ND$_3$}\xspace}
\newcommand{\HO}{\hbox{H$_2$O}\xspace}
\newcommand{\etal}{\emph{et al.}\xspace}
\newcommand{\PDEUN}{\ensuremath{P2_{1}2_{1}2_{1}}\xspace}

\newcommand{\micron}{\ensuremath{\mu}m\xspace}
\newcommand{\OLC}[1]{Ref.~[\onlinecite{#1}]\xspace}

\newcommand{\fcc}{\textit{fcc}\xspace}
\newcommand{\hcp}{\textit{hcp}\xspace}

\begin{document}
\title{Ionic ammonia ice}

\author{S. Ninet}
\email{sandra.ninet@impmc.upmc.fr}
\affiliation{Institut de Minéralogie et de Physique des Milieux
  Condensés, Université Pierre et Marie Curie - Paris 6, CNRS UMR
  7590, Paris, France}

\author{F. Datchi}
\affiliation{Institut de Minéralogie et de Physique des Milieux
  Condensés, Université Pierre et Marie Curie - Paris 6, CNRS UMR
  7590, Paris, France}

\author{P. Dumas}
\affiliation{Synchrotron SOLEIL, BP 48, 91192 Gif Sur Yvette, France.}

\author{M. Mezouar}
\affiliation{European Synchrotron Radiation Facility, BP 2220, F-38043
  Grenoble Cedex, France}

\author{G. Garbarino}
\affiliation{European Synchrotron Radiation Facility, BP 2220, F-38043
  Grenoble Cedex, France}

\author{A. Mafety}
\affiliation{Institut de Minéralogie et de Physique des Milieux
  Condensés, Université Pierre et Marie Curie - Paris 6, CNRS UMR
  7590, Paris, France}

\author{C.J. Pickard}
\affiliation{Department of Physics and Astronomy, University College
  London, Gower Street, London WC1E 6BT, United Kingdom}

\author{R.J. Needs}
\affiliation{Theory of Condensed Matter Group, Cavendish Laboratory, J
  J Thomson Avenue, Cambridge CB3 0HE, United Kingdom}

\author{A.M. Saitta}
\affiliation{Institut de Minéralogie et de Physique des Milieux
  Condensés, Université Pierre et Marie Curie - Paris 6, CNRS UMR
  7590, Paris, France}

\date{\today}


\begin{abstract}
  We report experimental and theoretical evidence that solid molecular
  ammonia becomes unstable at room temperature and high pressures and
  transforms into an ionic crystalline form. This material has been
  characterised in both hydrogenated (\NH) and deuterated (\ND)
  ammonia samples up to about 180 and 200 GPa, respectively, by
  infrared absorption, Raman spectroscopy and x-ray diffraction. The
  presence of a new, strong IR absorption band centered at 2500 \invcm
  in \NH (1900 \invcm in \ND) signals the ionization of ammonia
  molecules into \NHmoins and \NHplus ions, in line with previous
  theoretical predictions. We find experimental evidence for a
  coexistence of two crystalline ionic forms, which our \textit{ab
    initio} structure searches predict to be the most stable at the
  relevant pressures. The ionic crystalline form of ammonia is stable
  at low temperatures, which contrasts with the behaviour of water in
  which no equivalent crystalline ionic phase has been found.
\end{abstract}

\maketitle
\section{Introduction}

\begin{figure}
\includegraphics[width=3.3in]{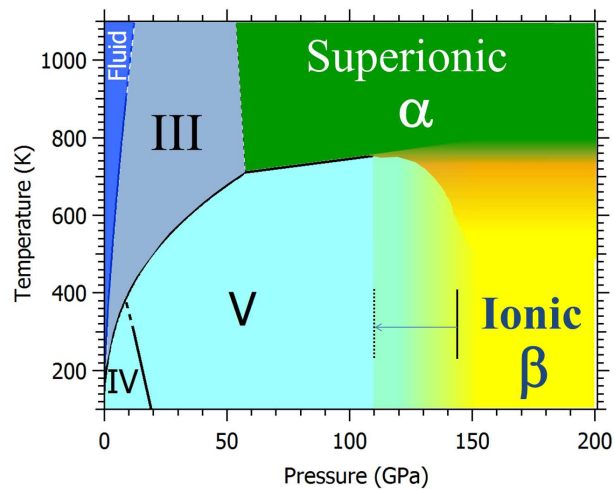}
\caption{\label{fig0} (color online). Phase diagram of ammonia. The
  blue, blue-grey and cyan coloured regions correspond to different
  molecular phases composed of \NH units. The yellow region represents
  the tentative stability domain of the ionic $\beta$ phase, composed
  of \NHplus and \NHmoins species, reported in the present work. The
  solid and dashed lines indicate the upstroke and downstroke
  transition pressures. At high pressures and temperatures, a dynamic
  ionisation is observed in the superionic $\alpha$ phase
  \cite{Ninet2012}, which is mainly composed of \NH, \NHplus and
  \NHmoins species.}
\end{figure}

The properties of ammonia at high pressures and temperatures are
important in planetary science and chemistry under extreme conditions.
Ammonia has a significant cosmic abundance and it is believed to be a
major constituent of the mantles of gas giant planets such as Neptune
and Uranus, and numerous extra solar (exo) planets. The high P-T phase
diagrams of ammonia, shown in Fig.~\ref{fig0}, and of water have been
the subject of recent investigations
\cite{Ninet2008,Ninet2012,Ojwang2012,Bethkenhagen2013,Goncharov2005c,Lin2005},
leading to the experimental discovery of a ``superionic'' phase
\cite{Ninet2012,Goncharov2005c}, as previously suggested by \textit{ab
  initio} calculations \cite{Cavazzoni1999}. The latter is
characterised by rapid diffusion of protons through the crystalline
nitrogen or oxygen lattice and could be relevant for understanding the
magnetic fields of giant icy (exo)planets.

The current study concerns the properties of ammonia at high pressures
but at lower temperatures than required for stabilising the superionic
phase. It has long been thought that solid ammonia would evolve
similarly to the isoelectronic water ice with pressure. However, the
hydrogen bonds in crystalline ammonia are weaker and more distorted
than in water ice, and while the latter evolves continuously towards a
symmetric H-bonded solid at around 100 GPa, where protons sit mid-way
between two oxygen atoms, no sign of H-bond symmetrisation in ammonia
has been found in \textit{ab initio} calculations up to 300 GPa
\cite{Cavazzoni1999,Fortes2003,Pickard2008}, or in experiments up to
120 GPa \cite{Sakashita1998,Datchi2006,Ninet2009}. Nonetheless, recent
work has suggested that the molecular ammonia solid is
thermodynamically unstable above about 100 GPa at low temperatures
\cite{Pickard2008}, and that it transforms into a crystalline ionic
solid phase with a structure of \textit{Pma2} symmetry consisting of
alternate layers of NH$_{4}^{+}$ and NH$_{2}^{-}$ ions. This ionic
solid has not been observed in experiments so far.

We present here a joint experimental and theoretical investigation
that provides strong evidences for the existence of an ionic phase of
solid ammonia at high pressures. \NH and \ND samples have been probed
at ambient temperature up to 194 and 184 GPa, respectively, using
Raman scattering, infrared (IR) absorption spectroscopy, and x-ray
diffraction techniques. In parallel, \textit{ab initio} theoretical
calculations, including new random structure searches, have been
carried out in the same pressure range.

\section{Experimental and theoretical methods}

Liquid \NH(99.99~\%, Air liquide) and \ND (99.96~\%, Eurisotop)
samples were loaded in membrane diamond anvil cells at 5 bar and 278
K. We used synthetic type IIas diamond anvils (Almax industries) with
flat culets of diameter 75 or 50 $\mu$m. The gaskets were made of
rhenium. The pressure was determined up to 70 GPa from the wavelength
shift of ruby fluorescence using the ruby scale from Dewaele \etal
\cite{Dewaele2004}, and at higher pressures, from the frequency shift
of the first-order Raman band of the diamond anvil tip
\cite{Loubeyre2002}.

\begin{figure}
\includegraphics[width=8cm]{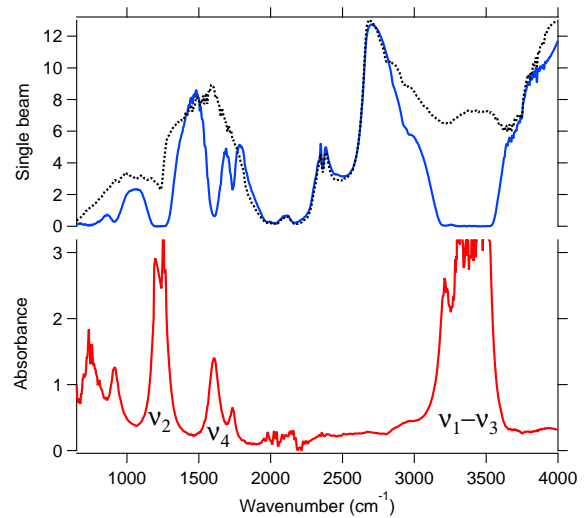}
\caption{\label{IRspectra_40GPa} (color online) Example of an IR
  spectrum collected from the molecular phase V (\PDEUN) of \NH sample
  at 40 GPa. The upper graph shows the transmission (single beam)
  spectr collected from \NH at 40 GPa (blue solid curve) and from a
  N$_2$ sample at 43 GPa (black dotted curve). The latter is used as
  the reference spectrum to calculate the absorbance of the \NH
  sample, which is displayed in the lower graph (red solid curve). The
  internal modes of phase V are indicated: $\nu_2$ and $\nu_4$ are
  respectively the symmetric and antisymmetric bending modes, $\nu_1$
  and $\nu_3$ are respectively the symmetric and antisymmetric
  stretching modes. The strong absorbance of $\nu_2$ and $\nu_3$
  saturates the absorption. The modes below 1000 \invcm are lattice
  modes.}
\end{figure}

IR absorption experiments were performed on the SMIS beamline of the
SOLEIL synchrotron facility (Saint-Aubin, France) using the FT-IR
spectrometer coupled to a vertical microscope. The IR beam was
condensed by 15$\times$ Cassegrain objectives to a spot of about 25
\micron in the focal plane. The sample absorbance is defined as
$A=-\log(I/I_0)$, where $I$ is the measured transmitted intensity and
$I_0$ the intensity of incident light. To correct for the absorption
of the diamond anvils and gasket, we used as $I_0$ the transmission
spectra measured in a separate experiment with similar diamond anvils
and gasket dimensions but using N$_2$ as sample, which is transparent
in the frequency range of interest.  Figure \ref{IRspectra_40GPa}
shows an example of IR spectra collected from \NH and N$_2$ at the
respective pressures of 40 and 43 GPa, and the calculated absorbance
of the \NH sample.

Raman spectra were collected at each pressure step using a DXR Raman
spectrometer from Thermo Fisher with 532 nm exciting laser
radiation. Additional Raman spectra were collected with an in-house
spectrometer using the 514.5 nm line of an argon laser.

X-ray diffraction experiments were conducted at beamline ID27 of the
European Synchrotron Radiation Facility (ESRF, Grenoble, France). We
used the angular-dispersive technique with monochromatic x-rays
($\lambda=0.3738$ \AA) and a bidimensional CCD detector (marCCD). The
x-ray beam was focussed to a spot of 2$\times$2.6 $\mu$m$^2$ FWHM on
the sample. Integration of the x-ray images was performed with the
\textsc{fit2D} software \cite{Hammersley1996}. Profile refinements
were conducted with the \textsc{Fullprof} software \cite{fullprof}.

Crystal structure prediction was carried out using the AIRSS
method~\cite{Pickard2011} and the CASTEP code\cite{CASTEP}, with four or eight ammonia formula units,
and a target pressure of 125 GPa. Except for the larger number of
molecules per unit cell, the same method as in
Ref.~\onlinecite{Pickard2008} was used.  For each phase considered, we
calculated the enthalpy/pressure/volume equations of state (EOS) up to
400 GPa, using the PWSCF code from the Quantum-Espresso
distribution~\cite{espresso}. These calculations were performed within
a standard Density-Functional-Theory framework, using the
Perdew-Burke-Ernzerhof gradient corrected functional \cite{PBE},
ultrasoft pseudopotentials \cite{Vanderbilt1990}, and a plane-wave
basis set.  A kinetic energy cutoff of 50 Ry and k-point grids
typically containing 16 to 32 points were used. Other functionals were
also tested to check the effect on the molecular-ionic transition
pressure, as explained in the next section. The vibrational modes and
corresponding Raman and IR spectra were calculated using the PHONON
code from the Quantum-Espresso package.

\section{Results and discussion}

\begin{figure}
\includegraphics[width=3.3in]{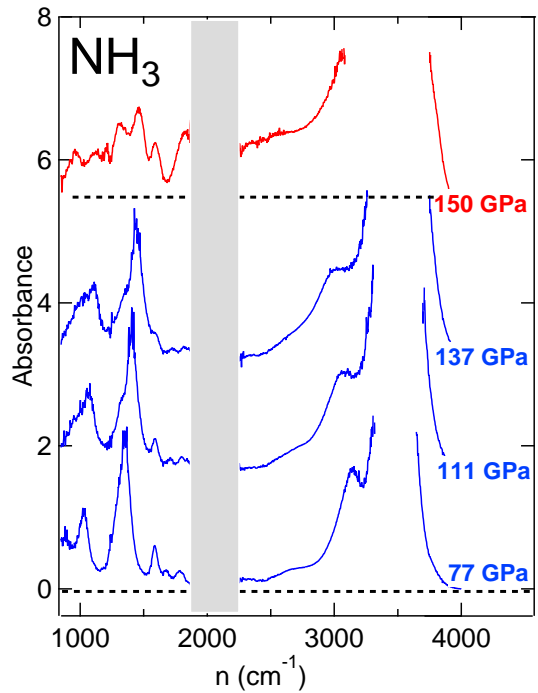}
\caption{\label{IRNH3} (color online) Evolution of the infrared
  absorption spectra of solid \NH with pressure. The spectra have been
  offset for easier visualization. The blue curves are for the
  molecular phase V, and the red one is for the ionic $\beta$ phase.
  The frequency window from $\sim$ 2000 to $\sim$ 2300 \invcm is
  obscured by the strong absorption band of the diamond
  anvils. Pressures are indicated on the right. }
\end{figure}

\begin{figure}
\includegraphics[width=3.3in]{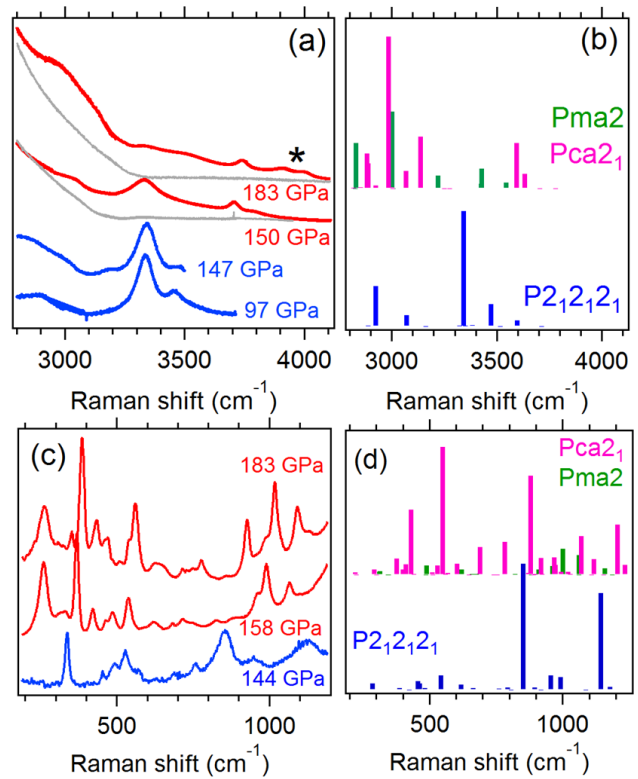}
\caption{\label{RamanNH3} (color online) Experimental [(a) and (c),
  300 K] and theoretical [(b) and (d), 0 K] Raman spectra of \NH are
  shown across the phase transition between the molecular phase V
  (\PDEUN) and the ionic $\beta$ phase. In (a) and (c), the blue and
  red curves show the experimental spectra, respectively, below and
  above the transition at the indicated pressures. The grey curves are
  spectra recorded with the laser spot focussed onto the Re gasket,
  showing the contribution of the second-order Raman band of the
  diamond anvils. In (b) and (d), the blue, pink and green bars show
  the DFPT-predicted Raman modes of, respectively, the \PDEUN phase V
  at 150 GPa, ionic $Pca2_{1}$ and $Pma2$ structures at 200
  GPa. Global offsets of $-100$ \invcm and $-60$ \invcm have been applied
  to the theoretical spectra in (b) and (d), respectively. The stars
  in (a) indicate two peaks which appeared after exposure to hard
  x-rays (33 keV), which were identified as the stretching vibrations
  of H$_{2}$ in a matrix of ammonia. We interpret this as a partial
  x-ray induced decomposition, as observed previously in H$_2$O at
  lower pressures \cite{Mao2006}. No peaks arising from molecular
  N$_{2}$ could be detected, which is not surprising since solid N$_2$
  forms a non-molecular amorphous state with no Raman-active modes at
  this pressure \cite{Gregoryanz2001}.}
\end{figure}

\begin{figure}
\includegraphics[width=3.3in]{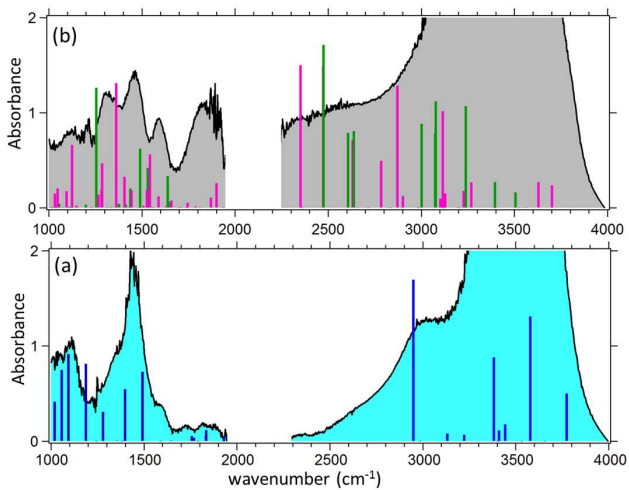}
\caption{\label{IRNH3_comp} (color online) The experimental mid-IR
  absorption spectra of ammonia are shown (a) at 144 GPa and (b) 158
  GPa, respectively, below and above the molecular-ionic phase
  transition at 300 K (black solid curves). The blue, green and pink
  bars show the theoretically predicted IR modes at 150~GPa and 0 K of
  the molecular \PDEUN, ionic \textit{Pma2}, and \textit{Pca2$_1$}
  structures, respectively. In order to make the lower frequency bands
  more visible, the theoretical IR intensities have been multiplied by
  6 for mode frequencies below 2000 \invcm.}
\end{figure}

\begin{figure*}
\includegraphics[width=6.5in]{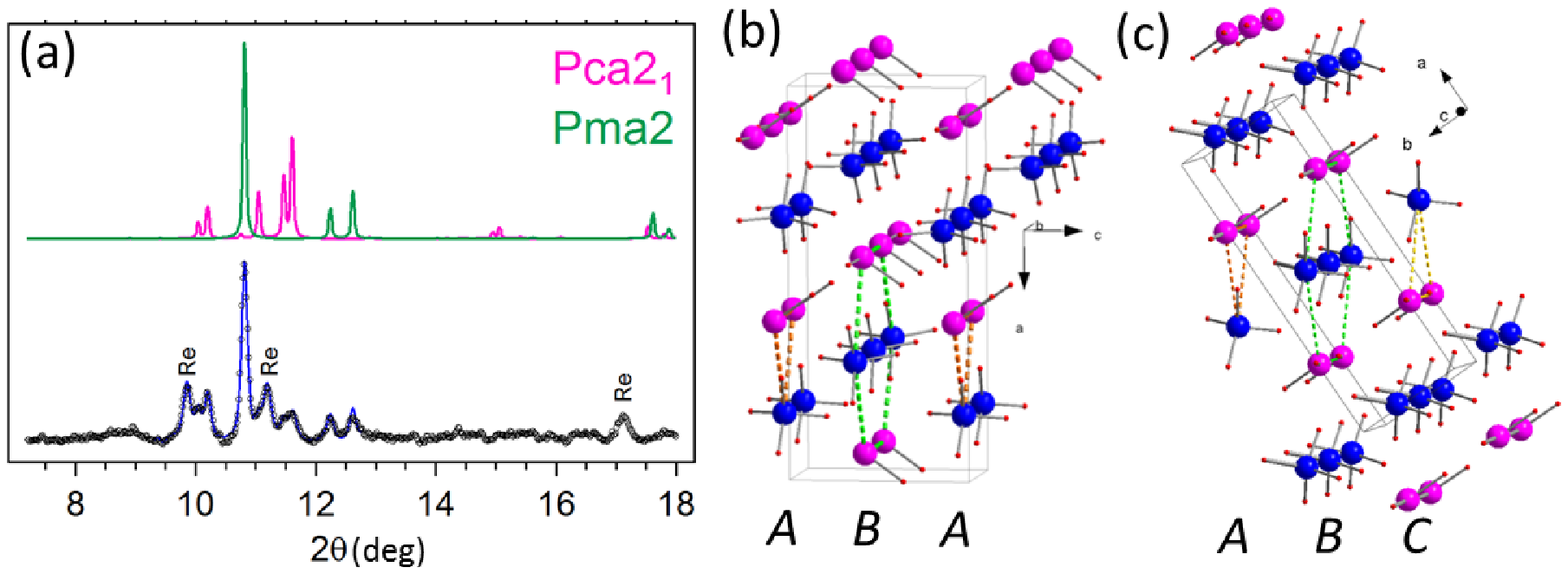}
\caption{\label{XrayNH3} (color online) X-ray pattern and structures
  of ionic ammonia. (a) The experimental (background-subtracted) x-ray
  pattern from the \NH sample at 194 GPa and 300 K is shown as black
  circles in the lower panel. The blue solid line is a Le Bail
  refinement obtained by considering a mixture of three phases:
  $Pma2$, $Pca2_1$ and Rhenium from the gasket (peaks indicated by
  ``Re''). In the upper part, the simulated x-ray powder patterns of
  the $Pca2_{1}$ (pink) and $Pma2$ (green) structures are depicted,
  using the cell parameters determined from the Le Bail refinement:
  $Pma2$: $a$ = 7.011 {\AA}, $b$ = 2.408 {\AA}, $c$ = 2.404 {\AA};
  $Pca2_1$: $a$ = 8.540 {\AA}, $b$ = 2.415 {\AA}, $c$ = 3.883
  {\AA}. (b) and (c): Representations of the ionic $Pca2_{1}$ and
  $Pma2$ structures, respectively. Red, blue and pink balls are,
  respectively, H atoms and the N atoms of the \NHmoins and \NHplus
  species. The pseudo \hcp (ABA) or \fcc (ABC) stacking of the
  nitrogen lattices are emphasised. The predicted cell parameters at
  200 GPa for $Pma2$ are: $a$=6.9756 \AA, $b$=2.371 \AA, $c$=2.370
  \AA, and for $Pca2_{1}$: $a$=8.471 \AA, $b$=2.380 \AA, $c$=2.863
  \AA. The differences between observed and predicted cell parameters
  are below 1.5 \%.}
\end{figure*}

Experimental IR and Raman spectra collected on compression of the \NH
samples are shown in Figs.~\ref{IRNH3} and \ref{RamanNH3},
respectively. Similar graphs for \ND are reported in Figs. 1 and 2 of
the Supplemental Material \cite{SupMat}. The vibrational spectrum of
the molecular \NH-V phase is complex, with 45 Raman and 35 IR
non-degenerate active modes~\cite{Ninet2006}. The strongly absorbing
IR bands corresponding to the symmetric bending ($\nu_2\sim1450$
\invcm at 137 GPa) and stretching ($\nu_1-\nu_3\sim 3000-3500$ \invcm)
vibrational modes saturate the absorption even though the sample
thickness is very small at megabar pressures (5-8 $\mu$m). By
contrast, the antisymmetric bending IR bands ($\nu_4$) and Raman
active lattice and molecular modes of \NH-V are clearly observed up to
150 GPa. The pressure dependencies of the measured Raman and IR mode
frequencies are presented in Figs. 3-5 of the Supplemental Material
\cite{SupMat}.

Above 150 GPa, strong changes in the experimental IR and Raman spectra
are observed, indicating a transition to another high-pressure phase
that we refer to as $\beta$-\NH. The latter is characterised by a new
strong and broad IR absorption band around 2300-2800 \invcm. The bands
in the frequency range corresponding to the bending modes
$\nu_2-\nu_4$ (from $\sim1200$ to $\sim1800$ \invcm) are also very
different from those of phase V. The Raman spectrum exhibits a new
peak in the antisymmetric stretching frequency range (around 3700
\invcm), a large decrease in intensity of the strongest peak of phase
V (around 3300 \invcm) and the appearance of a broad band around 3000
\invcm. Several new lattice Raman peaks also appear at low
frequencies, indicating that the new phase has a larger unit cell than
phase V (Fig.~\ref{RamanNH3}c). Similar changes are observed in \ND
above 150 GPa: the new IR bands in \ND appear around 1500 \invcm (non
saturated) and 1800-2000 cm$^{-1}$ (saturated), a new stretching Raman
peak appears at $\sim 2750$ \invcm, and a broad Raman band is observed
around 2000 \invcm. IR and Raman data were also collected on
decompression, the $\beta$ phase was then observed down to 110 GPa
before phase V was recovered. The large hysteresis (40 GPa) between
the transitions on the upstroke and downstroke suggests a substantial
kinetic barrier between the two phases, and therefore a likely change
in the molecular bonding at the transition.

The theoretical Raman and IR spectra predicted by density functional
perturbation theory (DFPT) for the molecular \PDEUN and the $Pma2$
ionic structures~\cite{Pickard2008} are displayed in Figs.\
\ref{RamanNH3} and \ref{IRNH3_comp}, respectively, where they can be
compared with the experimental ones. As previously noted for pressures
below 10 GPa~\cite{Ninet2006}, DFPT reproduces the experimental Raman
mode frequencies of phase V to better than 12 \%. The comparison is
more difficult for the experimental IR spectra because of the overlap
of bands and saturation in the spectra. The relative intensities of
the theoretical spectra are only in qualitative agreement with
experiment. We note that anharmonic effects are not taken into account
in the theoretical calculations, which may explain the discrepancies
with the experimental data.

The most noticeable difference between the calculated vibrational
properties of the $Pma2$ ionic structure and the \PDEUN molecular
phase is the occurrence in the former of intense IR modes in the
region 2460-2700 \invcm at 150 GPa, originating from the vibrations of
the \NHplus ions~\cite{Pickard2008}. This theoretical prediction gives
a good match to our observation of strong IR absorption around 2500
\invcm in the $\beta$-phase. The predicted IR bands of $Pma2$ around
1500-1650 \invcm also agree well with experiment; however there is no
strong IR absorption predicted for $Pma2$ around 1900 \invcm, in
contrast to observations for the $\beta$ phase, and the predicted
absorption in the range 3000-3700 \invcm is smaller than observed.

In the Raman spectra, the observed decrease in the intensity of the
main antisymmetric stretching mode (3400 \invcm) and the increase in
intensity around 3000 \invcm are consistent with the predicted Raman
features of the $Pma2$ phase. Nevertheless, the newly observed
stretching band at about 3700 \invcm, as well as the new lattice modes
below 300 \invcm, are absent in the $Pma2$ structure, which suggests
that this phase alone cannot account fully for the experimental
results.

As seen in Fig.~\ref{XrayNH3}a, there are also some inconsistencies
between the measured x-ray diffraction pattern of the $\beta$ phase at
194 GPa and the calculated one for the $Pma2$ structure. As a matter
of fact, although the reflections of the $Pma2$ structure are present
in the experimental pattern, there are several other peaks which it
cannot account for. These observations suggest either a larger unit
cell, a lower symmetry, or the coexistence of different structures.

\begin{figure}[t!]
\includegraphics[width=\columnwidth]{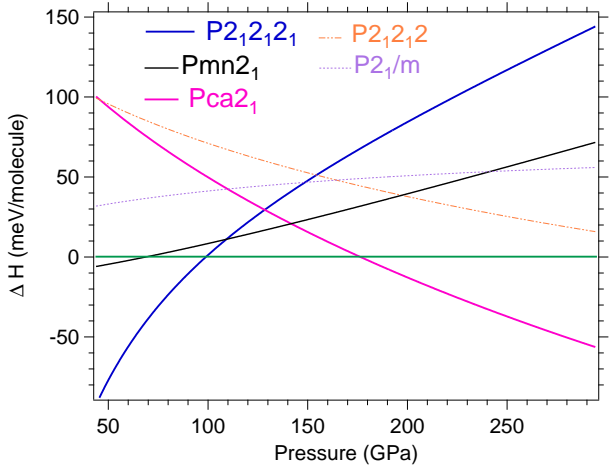}
\caption{\label{EnthalpyPlot} Difference in enthalpy with respect to
  the ionic $Pma2$ ($Z=4$) structure as a function of pressure.  The
  \PDEUN structure is that of the molecular phase V. The $Pma2$ ionic
  structure is identical to that reported in Ref.\
  \onlinecite{Pickard2008}. The other structures were obtained from
  new structural searches with up to $Z=8$ formula units per cell. }
\end{figure}

New \textit{ab-initio} random structural searches \cite{Pickard2011}
were therefore undertaken to determine whether other \NH structures
could exist in the pressure range of our experimental
observations. The same method as in Ref.~\onlinecite{Pickard2008} was
used, but the maximum number of \NH units per unit cell ($Z$) was
doubled with respect to the previous work, which used up to 4
molecular units. These searches found the previously reported
\cite{Pickard2008} $Pma2$ ($Z=4$) ionic structure and a number of new
structures, including an ionic structure of space group $Pca2_{1}$
($Z=8$). An enthalpy-pressure plot for the most competitive structures
is shown in Fig.\ \ref{EnthalpyPlot}. The $Pma2$ ionic structure is
confirmed to be the most stable between 100 and 176 GPa, while the
$Pca2_{1}$ ionic structure becomes thermodynamically preferred between
176 and 300 GPa. We have checked whether the appearance of a stable
ionic phase is robust to the choice of density functional, finding
that each functional tested predicts a transition from the molecular
phase V structure ($P2_12_12_1$) to the ionic $Pma2$ structure. The
transition pressure however varies with the functional chosen, ranging
from 68 GPa with LDA to 116 GPa with PBE0 \cite{PBE0}. Table
\ref{Tab:Ptrans_vs_func} lists the transition pressures for all cases
tested, resulting from previous \cite{Pickard2008, Griffiths2012} or
the present calculations.

Like $Pma2$, $Pca2_{1}$ is composed of \NHplus and \NHmoins ions, but
whereas $Pma2$ consists of alternate layers of \NHplus and \NHmoins
and has a pseudo face-centered cubic (\fcc) N-lattice, $Pca2_1$ is
composed of zig-zag chains with a pseudo hexagonal close-packed (\hcp)
N-lattice (Fig.~\ref{XrayNH3}b). Another ionic structure of space
group $P2_12_12$ ($Z=8$) was also found, which differs from $Pca2_1$
only in the orientation of the \NHmoins ions, but this phase is never
stable in the pressure range studied (see Fig.\
\ref{EnthalpyPlot}). Other, very different structures were found: the
``mixed'' molecular-ionic $Pmn2_1$ structure, composed of \NHplus,
\NHmoins ions and \NH molecules, and the $P2_1/m$ structure, composed
of N$^{3-}$ and \NHplus ions. Although unstable at 0 K, these
structures are close in enthalpy to the $Pma2$ and $Pca2_1$ phases,
especially the mixed $Pmn2_1$ structure, which remains within 25
meV/molecule of the $Pma2$ structure up to $\sim$160 GPa (see Fig.\
\ref{EnthalpyPlot}). The structural parameters, representations and
simulated x-ray patterns of all of the phases discussed above are
given respectively in Table 1 and Figs.\ 6 and 7 of the Supplemental
Material \cite{SupMat}.

As shown in Fig.~\ref{XrayNH3}a, the experimental x-ray pattern at 194
GPa can be interpreted as arising from the coexistence of the fully
ionic $Pma2$ and $Pca2_{1}$ structures: the fitted unit cell
parameters are within 1.5 \% of the theoretical ones (see caption of
Fig.~\ref{XrayNH3}). This coexistence is reasonable as only a small
energy separates the two ionic phases over a substantial pressure
range around 170 GPa. The x-ray pattern presents too much texture
(preferred orientation) to allow the determination of the relative
amounts of the two phases from full-profile refinement. Regarding the
Raman and IR spectra, the main signature of the ionic species in the
$Pca2_{1}$ structure is, as in $Pma2$, the intense IR modes in the
frequency range 2300-2800 cm$^{-1}$, where the molecular phases have
no modes at all. In addition, the $Pca2_1$ structure shows strong IR
activity around 1900 \invcm and 3700 \invcm, and two Raman modes
around 3600 \invcm, in reasonable agreement with the experimental
observations.  A coexistence of the two ionic structures therefore
matches quantitatively the x-ray data, and qualitatively the Raman and
IR data. Clearly, the agreement between the observed and theoretical
intensities for the Raman and IR spectra is poor, but as observed
above for the molecular \PDEUN phase, this likely originates from the
approximations used in the theoretical approach, which is limited to
the harmonic level.

We also considered the possibility of a coexistence of the ionic
$Pma2$ and mixed $Pmn2_1$ structures. Since $Pmn2_1$ is partially
ionic, its Raman and IR spectra have similar features as the $Pma2$
and $Pca2_1$ structures (see Fig.\ 8 of the Supplemental Material
\cite{SupMat}). The Bragg reflections which are not indexed by $Pma2$
or the gasket can also be well indexed by $Pmn2_1$ (see Fig.\ 9 of the
Supplemental Material \cite{SupMat}); however, the ratios of the
refined cell parameters for the $Pmn2_1$ structure are $b/a=1.103$ and
$c/a=1.248$, which are quite different from the theoretical ones:
$b/a=1.046$ and $c/a=1.221$. Since, according to our calculations, the
$Pmn2_1$ structure does not have a pressure range of thermodynamic
stability, this coexistence model thus appears less likely. The same
arguments also militate against coexistence of the molecular \PDEUN
phase with $Pma2$ at this pressure. Indeed, the theoretical $b/a$ and
$c/a$ ratios for \PDEUN at 200~GPa are 1.736 and 1.631,
respectively. These agree well with the extrapolation of experimental
data \cite{Datchi2006} up to 120~GPa, and are largely different from
the cell axis ratios which are obtained from our x-ray pattern at 194
GPa. Moreover, the enthalpy difference between \PDEUN and $Pca2_1$
increases rapidly with pressure and exceeds 0.1~eV/molecule at
200~GPa.

Of course, we cannot exclude the possibility that another structure
with a larger unit cell -- that could result from a more complex
stacking of the ionic planes, might explain the experimental data. We
emphasise, however, that such a structure \emph{must be at least
  partially ionic} to explain the infrared absorption spectrum of the
$\beta$-phase. Indeed, whatever the crystallographic structure, the IR
band around 2500 \invcm (1900 \invcm in \ND) in the $\beta$ phase
cannot be due to molecular \NH, which therefore implies the presence
of \NHplus ions.

It is interesting to note that all of the crystalline phases of
ammonia have a nitrogen sub-lattice that can be described as pure or
pseudo-\hcp or \fcc. This is the case for the proton-disordered
molecular phases II (\hcp) and III (\fcc), and the proton-ordered
molecular phases I (pseudo-\fcc), IV and V (pseudo-\hcp). Even the
high P-T superionic $\alpha$ phase may have either a \fcc or a
pseudo-\hcp N lattice, depending on whether it is obtained by
compressing phase III or annealing phase V, respectively
\cite{Ninet2012}. Our findings extend this property to the
low-temperature ionic phases $Pma2$ (pseudo-\fcc) and $Pca2_1$
(pseudo-\hcp). Since the N atoms already adopt densely packed
configurations and are clearly reluctant to undergo a rearrangement,
the ``job'' of minimising the energy as density increases is
accomplished by rearrangements of the H atoms. As pointed out in Ref.\
\onlinecite{Pickard2008}, the volumes per formula unit of the ionic
phases are much smaller than in the molecular ones, because the shapes
of the \NHplus and \NHmoins ions enable them to pack more tightly than
\NH molecules. The ionic structures are therefore favoured at high
density. We also note that the $Pca2_1$ structure strongly resembles
the $P2_{1}/m$ ionic phase predicted to be stable above 300 GPa
\cite{Pickard2008}, where \NHmoins ions are linked in zig-zag chains
by symmetric H-bonds.

\begin{table}[t]
\center
\caption{\label{Tab:Ptrans_vs_func} Predicted transition pressures between
  the molecular \PDEUN and ionic $Pma2$ structures according to different
  DFT functional: PBE \cite{PBE}, PBE + Tkatchenko-Scheffler van der Waals
  (TS-vdW) scheme \cite{TS-vdW}, LDA, HSE06 \cite{HSE06} and PBE0 \cite{PBE0}.
  The results of calculations with the PBE, LDA and PBE0 functionals were previously
  reported in \OLC{Pickard2008} and \OLC{Griffiths2012}, those for the PBE+TS-vdW
  and HSE06 functionals were performed in the present work. The molecular-ionic
  transition is predicted in every case.}
\begin{ruledtabular}
\begin{tabular}{cccccc}
 Functional &   PBE   & PBE+TS-vdW &  LDA   &    HSE06  &    PBE0 \\

 P (GPa) &     97  &      83    &    68    &    126  &      116 \\
\end{tabular}
\end{ruledtabular}
\end{table}

\section{Conclusion}

In summary, our experimental investigation of solid ammonia to
pressures of nearly 200~GPa shows that a transition to a previously
unobserved phase, denoted $\beta$, occurs at 150 GPa on compression,
which remains stable down to 110 GPa on decompression. The important
changes in the Raman and IR active internal modes with respect to the
molecular phase V indicate a drastic modification of the
intramolecular bonding across the transition. The appearance of
intense IR bands in the frequency range around 2500 \invcm makes a
strong case in favour of an ionic structure for the $\beta$ phase,
composed of \NHplus and \NHmoins ions, as predicted in a previous
theoretical work \cite{Pickard2008}. The latter has been expanded in
the present study, with two main results: first, the predicted
transition between the molecular \PDEUN and the ionic $Pma2$ around
100 GPa is proved to be robust against the choice of density
functional. Second, our new structural searches with up to 8 formula
units per unit cell demonstrate that $Pma2$ remains the most
competitive phase above 100 GPa at 0 K, and that another ionic
structure with space group $Pca2_1$ ($Z=8$), becomes more stable above
176 GPa. The experimental Raman, IR and x-ray data are best explained
by a coexistence of the $Pma2$ and $Pca2_1$ ionic structures near 190
GPa, which is consistent with the small enthalpy difference ($\sim15$
meV/molecule) between the two phases. Although we cannot presently
exclude the possibility that the $\beta$ phase has a different
crystallographic structure, with perhaps a larger number of formula
units per unit cell, the presence of strong IR absorption in the range
of \NHplus vibrations shows that this structure is composed, at least
partially, of ionic species.

Our results confirm that, despite their resemblance at ambient and
moderate pressures, \NH and \HO follow different structural evolutions
at very high densities; while ammonia transforms from a molecular to
an ionic ammonium amide solid at megabar pressures, water ice evolves
into a symmetric H-bonded solid in the same pressure range. In Ref.\
\onlinecite{Pickard2008} it was pointed out that the cost of forming a
solid composed of H$_3$O$^+$ and OH$^-$ units is larger than for solid
ammonium amide (1.5 and 0.9 eV/unit respectively). This adds to the
fact that water ice can form compact structures accommodating strong
and symmetric H-bonds, whereas in ammonia ice the close-packed
N-lattice leads to weaker and inequivalent H-bonds that cannot
symmetrise without a large deformation of the molecules
\cite{Gauthier1988}. We also note that mixtures of ammonia and water
have been predicted to form ionic ammonium hydroxide solids under
pressure \cite{Griffiths2012}. It would be interesting to extend the
present work in this direction, especially since ammonia hydrates are
relevant for the description of icy planets and their moons. Finally,
further investigations are required to understand the relationship
between the static ionic solid found here and the dynamic superionic
one reported in Ref.~\onlinecite{Ninet2012} at high P-T conditions.

\begin{acknowledgments}
  We acknowledge the synchrotron SOLEIL for provision of beam time
  allocated to proposals 20090281 and 20110969 and the ESRF for
  in-house beam time at ID27. The computational part of this work was
  performed using HPC resources from GENCI/IDRIS (Grant 2012-091387).
  We thank Sylvain Petit Girard (ESRF-ID27) and Gunnar Weck (CEA) for
  their help with x-ray experiments, and Christophe Sandt and Frédéric
  Jamme (SOLEIL - SMIS) for their help with IR experiments. C.J.P.\
  and R.J.N.\ acknowledge financial support from the EPSRC.
\end{acknowledgments}


\end{document}